\documentclass[pre,twocolumn,aps,showpacs]{revtex4}
\newcommand{\bec}[1]{\mbox{\boldmath $ #1$}}
\usepackage{graphicx}
\begin{document}
\title{Turbulent Thermal Diffusion in a Multi-Fan Turbulence
Generator with the Imposed Mean Temperature Gradient}

\author{A.~Eidelman}
\author{T.~Elperin}
\email{elperin@bgu.ac.il}
\author{N.~Kleeorin}
\author{I.~Rogachevskii}
\author{I.~Sapir-Katiraie}
\affiliation{The Pearlstone Center for Aeronautical Engineering
Studies, Department of Mechanical Engineering, Ben-Gurion University
of the Negev, Beer-Sheva 84105, P. O. Box 653, Israel}
\date{\today}
\begin{abstract}
We studied experimentally the effect of turbulent thermal diffusion
in a multi-fan turbulence generator which produces a nearly
homogeneous and isotropic flow with a small mean velocity. Using
Particle Image Velocimetry and Image Processing techniques we showed
that in a turbulent flow with an imposed mean vertical temperature
gradient (stably stratified flow) particles accumulate in the
regions with the mean temperature minimum. These experiments
detected the effect of turbulent thermal diffusion in a multi-fan
turbulence generator for relatively high Reynolds numbers. The
experimental results are in compliance with the results of the
previous experimental studies of turbulent thermal diffusion in
oscillating grids turbulence (Buchholz et al. 2004; Eidelman et al.
2004). We demonstrated that turbulent thermal diffusion is an
universal phenomenon. It occurs independently of the method of
turbulence generation, and the qualitative behavior of particle
spatial distribution in these very different turbulent flows is
similar. Competition between turbulent fluxes caused by turbulent
thermal diffusion and turbulent diffusion determines the formation
of particle inhomogeneities.
\end{abstract}

\maketitle

\section{Introduction}

The main goal of this study is to describe the experimental
investigation of the effect of turbulent thermal diffusion in a
multi-fan turbulence generator. Turbulent thermal diffusion is
associated with the correlation between temperature and velocity
fluctuations in a turbulent flow with an imposed mean temperature
gradient and causes a relatively strong non-diffusive mean flux of
particles in the direction of the mean heat flux. This effect
results in the formation of large-scale inhomogeneities in
particle spatial distribution whereby particles accumulate in the
vicinity of the minimum of the mean fluid temperature. Turbulent
thermal diffusion was predicted theoretically by Elperin et al.
(1996; 1997) and detected experimentally by Buchholz et al. (2004)
and Eidelman et al. (2004) in oscillating grids turbulence.

The mechanism of the phenomenon of turbulent thermal diffusion for
inertial solid particles is as follows. Inertia causes particles
inside the turbulent eddies to drift out to the boundary regions
between the eddies (i.e., regions with low vorticity and maximum of
fluid pressure). Therefore, particles accumulate in regions with
maximum pressure of the turbulent fluid. Similarly, there is an
outflow of particles from regions with minimum pressure of fluid. In
a homogeneous and isotropic turbulence without large-scale external
gradients of temperature, a drift from regions with increased or
decreased concentration of particles by a turbulent flow of fluid is
equiprobable in all directions, as well as pressure and temperature
of the surrounding fluid do not correlate with the turbulent
velocity field. Therefore, only turbulent diffusion determines the
turbulent flux of particles.

In a turbulent fluid flow with a mean temperature gradient, the mean
heat flux is not zero, i.e., the fluctuations of temperature and the
velocity of the fluid are correlated. Fluctuations of temperature
cause fluctuations of fluid pressure. These fluctuations result in
fluctuations of the number density of particles. Indeed, an increase
of pressure of the surrounding fluid is accompanied by an
accumulation of particles due to their inertia. Therefore, the
direction of the mean flux of particles coincides with that of the
heat flux, and the mean flux of particles is directed to the region
with minimum mean temperature, and the particles accumulate in this
region (Elperin et al. 1996).

The mechanism of turbulent thermal diffusion is associated with a
nonzero divergence of a particle velocity field. The latter is
caused either by particle inertia or inhomogeneity of fluid density
in a non-isothermal low-Mach number turbulent fluid flow. Therefore,
the effect of turbulent thermal diffusion can be also observed in
the suspension of non-inertial particles (e.g., particles with a
Stokes time of the order of $10^{-6} - 10^{-5}$ s in air flows) or
for gaseous admixtures in non-isothermal low-Mach number turbulent
fluid flows (see Elperin et al., 1997). Note that when we refer to
compressible non-isothermal fluid flow with low Mach numbers, it
means that ${\rm div} \, (\rho \, {\bf v}) \approx 0$, where ${\bf
v}$ is the fluid velocity and $\rho$ is the fluid density. The
latter implies that ${\rm div} \, {\bf v} \approx - ({\bf v} \cdot
\bec{\nabla}) \rho / \rho \not = 0 .$ In particular, in a
non-isothermal fluid flow with a temperature gradient ${\rm div} \,
{\bf v} \approx - ({\bf v} \cdot \bec{\nabla}) \rho / \rho \approx
({\bf v} \cdot \bec{\nabla}) T / T \not = 0$, where $T$ is the fluid
temperature.

Numerical simulations, laboratory experiments and observations in
the atmospheric turbulence revealed formation of long-living
inhomogeneities in spatial distribution of small inertial particles
and droplets in turbulent fluid flows (see, e.g., Wang and Maxey
1993; Korolev and Mazin 1993; Eaton and Fessler 1994; Fessler et al.
1994; Maxey et al. 1996; Aliseda et al. 2002; Shaw 2003). The origin
of these inhomogeneities was intensively studied by Elperin et al.
(1996; 1997; 2000a; 2000b). It was pointed out that the effect of
turbulent thermal diffusion is important for understanding different
atmospheric phenomena (e.g., atmospheric aerosols, smog formation,
etc). In particular, the existence of a correlation between the
appearance of temperature inversions and the aerosol layers
(pollutants) in the vicinity of the temperature inversions is well
known (see, e.g., Csanady 1980; Seinfeld 1986; Flagan and Seinfeld
1988). Turbulent thermal diffusion can cause the formation of
large-scale aerosol layers in the vicinity of temperature inversions
in atmospheric turbulence (Elperin et al. 2000a; 2000b).
Observations of the vertical distributions of pollutants in the
atmosphere show that maximum concentrations can occur within
temperature inversion layers (see, e. g., Csanady 1980; Seinfeld
1986; Jaenicke 1987). The characteristic parameters of the
atmospheric turbulent boundary layer are: the maximum scale of
turbulent flow is $L \sim 10^3 - 10^4 $ cm; the turbulent fluid
velocity in the scale $L$ is $ u \sim 30 - 100 $ cm/s; the Reynolds
number is $ {\rm Re} = u L /\nu \sim 10^6 - 10^7$ (see, e. g.,
Csanady 1980; Seinfeld 1986; Blackadar 1997), where $\nu$ is the
kinematic viscosity. For instance, for particles with material
density $ \rho_p \sim 1 - 2 $ g / cm$^3 $ and radius $a = 20 \,
\mu$m the characteristic time of formation of inhomogeneities is of
the order of $ 1 $ hour for the temperature gradient $ 1 K / 100 $ m
and $ 2 $ hours for the temperature gradient $ 1 K / 200 $ m. The
effect of turbulent thermal diffusion might be also of relevance in
different industrial non-isothermal turbulent flows (Elperin et al.
1998).

Turbulent thermal diffusion is a new and fundamental phenomenon.
Therefore, this phenomenon should be studied for different types of
turbulence and different experimental set-ups. Previously phenomenon
of turbulent thermal diffusion was studied experimentally only in
oscillating grids turbulence whereby the Reynolds numbers were not
so high (see for details Buchholz et al. 2004; Eidelman et al.
2004). In order to study the effect of turbulent thermal diffusion
at higher Reynolds numbers we constructed a multi-fan turbulence
generator. Similar apparatus was used in the past in turbulent
combustion studies (Birouk et al. 1996) and in studies of
turbulence-induced preferential concentration of solid particles in
microgravity conditions (Fallon and Rogers 2002). The multi-fan
turbulence generator allows us to produce nearly homogeneous
isotropic turbulent fluid flow with a small mean velocity. Using
Particle Image Velocimetry and Image Processing Techniques we
determined velocities and spatial distribution of tracer particles
in the isothermal and non-isothermal turbulent flows. Our
experiments with non-isothermal turbulent flows were performed for a
stably stratified fluid flow with a vertical mean temperature
gradient which was formed by a cooled bottom wall and a heated top
wall of the chamber. We found that particles accumulate in the
vicinity of the bottom wall of the chamber (where the mean fluid
temperature is minimum) due to the effect of turbulent thermal
diffusion. Sedimentation of particles in a gravity field in these
experiments was very slow in comparison with accumulation of
particles caused by turbulent thermal diffusion. In the present
study we demonstrated that turbulent thermal diffusion occurs
independently of the method of turbulence generation.

\section{Experimental set-up}

A multi-fan turbulence generator includes eight fans (120 mm in
outer diameter and with controlled rotation frequency of up to 2800
rpm) mounted in the corners of a cubic Perspex box and facing the
center of the box. The Perspex box is a cube $400 \times 400 \times
400$ mm$^3$, with eight 272 mm equilateral triangles mounted in its
corners used as solid bases for the fans (see Fig.~1). Each fan was
calibrated separately, and the input current and rotation speed were
measured and logged. We also tested one fan operating above an
electric heat source. We placed a thermocouple in the vicinity of
the fan's motor, and repeated this test for different temperatures
(290, 310, 330 K). These experiments showed that the fan rotation
speed does not depend on the temperature.

\begin{figure}
\centering
\includegraphics[width=7cm]{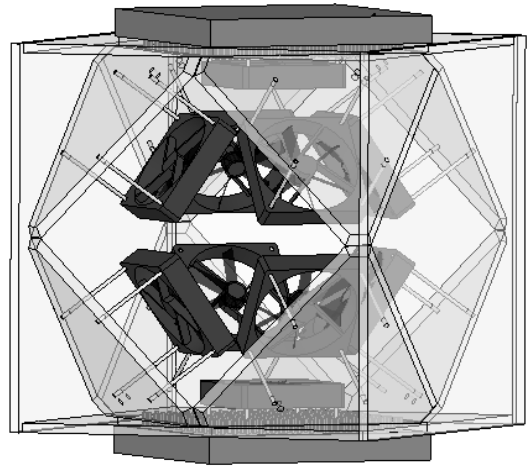}
\caption{\label{Fig1} The scheme of the test section.}
\end{figure}

At the top and bottom  walls of the Perspex box we installed two
heat exchangers with rectangular $3 \times 3 \times 15$ mm$^3$ fins.
The upper wall was heated up to 343 K, the bottom wall was cooled to
283 K. Therefore, comperatively large vertical mean temperature
gradient $(\sim 92$ K/m) was formed in the core of the flow. The
temperature was measured with a high-frequency response thermocouple
($0.005$ inches in diameter) which was glued externally to a wire
$(2.2$ mm of Cu in diameter coated with $0.85$ mm Teflon). The
accuracy of the temperature measurements was of the order of $0.1$
K. We found that the temperature measurements affected the flow in a
very small area around the wire. Two additional fans were installed
at the bottom and top walls of the chamber in order to produce a
large mean temperature gradient in the core of the flow. Our
measurements showed that these two additional fans only weakly
affected considerably the homogeneity and isotropy of the turbulent
flow.

Velocity fields were measured using Particle Image Velocimetry
(PIV) technique. The flow was seeded with incense smoke and was
illuminated by a Surelite LSI-10 (Continuum) Nd:YAG pulsed laser
with a power of 170 mJ/pulse. The light sheet optics includes
spherical and cylindrical Galilei telescopes with tuneable
divergence and adjustable focus length. We used a progressive-scan
12 bit digital CCD camera (pixel size $6.7 \, \mu$m $\times 6.7 \,
\mu$m each) with a dual-frame-technique for cross-correlation
processing of captured images. A programmable Timing Unit (PC
interface card) generated sequences of pulses to control the
laser, camera and data acquisition rate. The data was processed
using standard cross correlation techniques (DaVis 7.0 code,
LaVision, G\"{o}ttingen).

An incense smoke with sub-micron particles as a tracer for the PIV
measurements was produced by high temperature sublimation of solid
incense particles. Analysis of smoke particles using a microscope
(Nikon, Epiphot with an amplification of 560) and a PM-300 portable
laser particulate analyzer showed that these particles have an
approximately spherical shape and that their mean diameter is of the
order of $0.7 \mu$m. In order to prevent from any thermal effects
caused by the incense smoke generator, we placed it far away from
the test section behind a wall, so that the incense smoke was
transported through a five meter long pipeline and was cooled before
it entered the test section. The smoke was feeded into the test
section at room temperature. We measured the smoke temperature
inside the pipeline at two locations: 0.5 m and 3.5 m from the
generator. At the first location the smoke temperature was 318 K,
while at the second location it was 294 K. The number density of
smoke particles inserted to the test section in the experiments was
of the order of $10^4$ cm$^{-3}$.

We determined mean and r.m.s. velocities, two-point correlation
functions and an integral scale of turbulence from the measured
velocity fields. A series of 130 pairs of images acquired with a
frequency of 4 Hz were stored for calculating the velocity maps and
for ensemble and spatial averaging of turbulence characteristics.
The center of the measurement region coincides with the center of
the chamber. We measured the velocity in flow area of $92 \times 92$
mm$^2$ with a spatial resolution of $1024 \times 1024$ pixels. These
regions were analyzed with interrogation windows of $32 \times 32$
pixels. A velocity vector was determined in every interrogation
window, allowing us to construct a velocity map comprising $32
\times 32$ vectors. The mean and r.m.s. velocities for each point of
the velocity map (1024 points) were determined by averaging over 130
independent maps, and then over 1024 points. The two-point
correlation functions of the velocity field were determined for each
point of the central part of the velocity map ($16 \times 16$
vectors) by averaging over 130 independent velocity maps, and then
over 256 points. Our tests showed that 130 image pairs contain
enough data to obtain reliable statistical estimates. An integral
scale $L$ of turbulence was determined from the two-point
correlation functions of the velocity field. For this end we used
exponential approximation of the correlation function since the
experimentally measured correlation function did not reach zero
values.

\begin{figure}
\centering
\includegraphics[width=8cm]{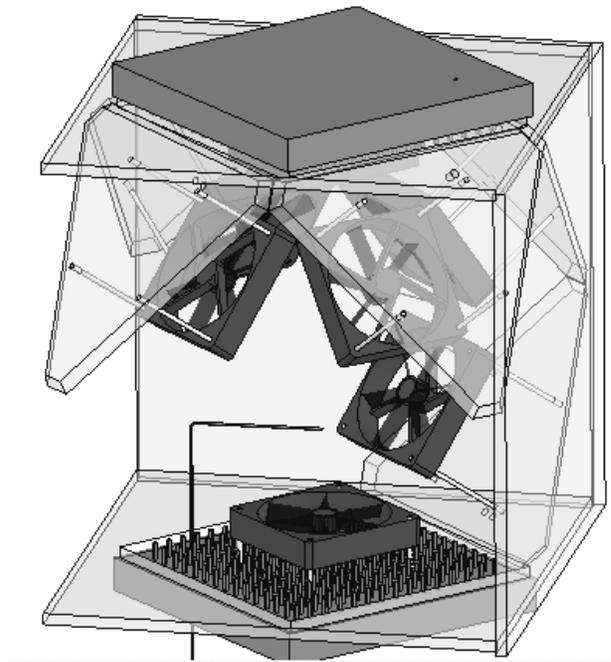}
\caption{\label{Fig2} The scheme of the temperature measurements.
The wire that holds the thermocouple is inserted from the bottom of
the chamber (some parts of the test section are not shown).}
\end{figure}

\begin{table}
\label{tab1}
\begin{tabular}{|l|c|c|}
\multicolumn{3}{c}{Table 1}\\
\multicolumn{3}{c}{Flow parameters in a multi-fan turbulence generator}\\
\hline
Horizontal and vertical directions    &       Y    &     Z \\
 \hline
Reynolds number ${\rm Re} = u \, L / \nu$    &      703   &    875
\\
 \hline
Integral length scale $L$ (mm)      &       14.85  &  16.4 \\
 \hline
r.m.s. velocity $u$  (m/s)         &         0.71  &   0.8 \\
 \hline
Turbulence integral time scale   $\tau =L/ u$ (ms)      &      20.9  &     20.5 \\
 \hline
Rate of dissipation $\varepsilon=u^3/ L$ (m$^2/$ s$^3$) & 24.1 &  31.2 \\
  \hline
Taylor microscale $\lambda = \sqrt{15 \, \nu \, \tau}$ (mm)    &       2.17  &  2.15 \\
 \hline
Kolmogorov length scale $\eta=L \, {\rm Re}^{-3/4}$ ($\mu$m)  &    109  &   102 \\
\hline
$Re_\lambda = u \, \lambda / \nu$ in the Taylor microscale       &   103   &    115 \\
 \hline

\end{tabular}
\end{table}

Spatial resolution of of the velocity measurements was about $2.9$
mm for a probed area $92 \times 92$ mm$^2$ with interrogation
window $32 \times 32$ pixels. The maximum tracer particle
displacement in the experiment was of the order of $8$ pixels,
i.e., $1/4$ of the interrogation window. The average displacement
of tracer particles was of the order of $2.5$ pixels. Therefore,
the average accuracy of the velocity measurements was of the order
of $4 \%$ for the accuracy of the correlation peak detection in
the interrogation window which was of the order of $0.1$ pixels
(see, e.g., Adrian 1991; Westerweel 1997; 2000).

In order to create large mean temperature gradient, the top and
bottom fans were run at different speeds than the peripheral fans
(see below). This regime was found empirically. Clearly, this
introduces a weak anisotropy in velocity fluctuations in the
vertical direction which is less than $10 \%$. Note that the
anisotropy of turbulence is not essential for the validation of the
phenomenon of turbulent thermal diffusion.

Fluid flow parameters (at the rotation speed $1500$ rpm for eight
corner fans and $2300$ rpm for top and bottom fans) at the
horizontal $Y$ and vertical $Z$ directions are presented in Table 1.
In the experiments the maximum mean flow velocity was of the order
of $0.1 - 0.2$ m/s while the r.m.s. velocity was of the order of
$0.7 - 1.1$ m/s. Thus, the measured r.m.s. velocity was much higher
than the mean fluid velocity in the core of the flow. Figure~3 shows
the two-point correlation functions of the velocity field at the
horizontal and vertical directions. In particular, in Fig.~3 we
plotted the longitudinal velocity correlation coefficients, $f(y) =
\langle u_y(0) u_y(y) \rangle / \langle u_y^2(0) \rangle $ and $f(z)
=\langle u_z(0) u_z(z) \rangle / \langle u_z^2(0) \rangle $, where
${\bf u}$ are the fluid velocity fluctuations. Figure~3 and Table 1
demonstrate that the multi-fan turbulence generator produced a
weakly anisotropic turbulent fluid flow with a small mean velocity.

The energy spectrum of the fluid flow is shown in Fig.~4. The
one-dimensional longitudinal  energy spectrum was determined by a
standard procedure. In particular, we determined the Fourier
components $u_y(k_y)$ and $u_z(k_z)$ of the fluctuating velocity
field, and then determined $\langle |u_y(k_y)|^2 \rangle$ and
$\langle |u_z(k_z)|^2 \rangle$, where $\langle ... \rangle$ is the
ensemble averaging over 130 independent velocity maps. In Fig.~4 we
plotted horizontal and vertical energy component spectra. As can be
seen from Fig.~4, the energy spectrum is different from $-5/3$ law.
However, our study showed that the existence of the effect of
turbulent thermal diffusion is independent of the slope of the
energy spectrum.

\begin{figure}
\centering
\includegraphics[width=8cm]{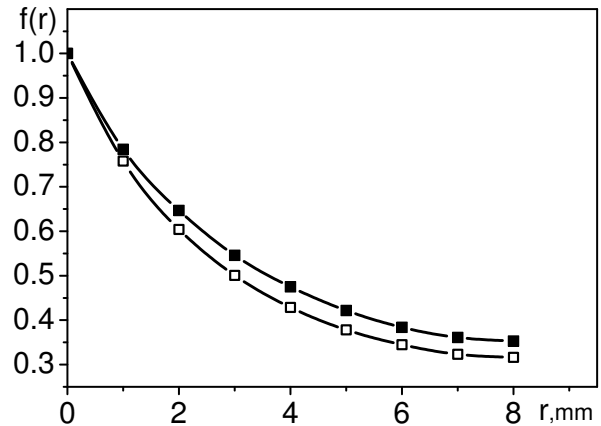}
\caption{\label{Fig3} The two-point longitudinal correlation
functions of the velocity field at the horizontal direction (filled
squares) and vertical direction (unfilled squares).}
\end{figure}

\begin{figure}
\centering
\includegraphics[width=8cm]{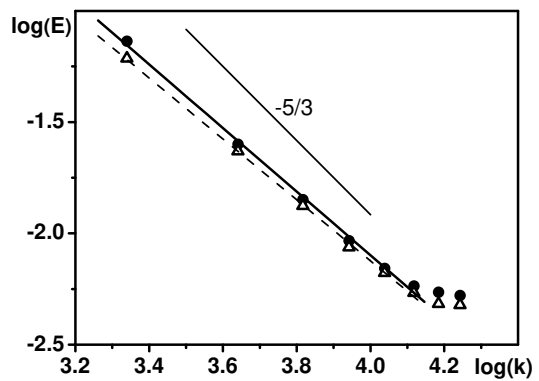}
\caption{\label{Fig4} The energy spectrum of the fluid flow.}
\end{figure}

Spatial particle number density distributions were obtained using a
single frame from the double frame captured for PIV measurements.
For this purpose the intensity of laser light Mie scattering by
tracer particles was recorded and averaged over 130 single frames.
We probed the central $ 9.2 \times 9.2 $ cm region in the chamber by
determining the mean intensity of scattered light in $ 32 \times 16
$ interrogation windows with a size of $ 32 \times 64 $ pixels. The
vertical distribution of the intensity of the scattered light was
determined in 16 vertical strips, which are composed of 32
interrogation windows. Variations of the obtained vertical
distributions between these strips were very small. We used spatial
average across the strips and ensemble average over 130 images of
the vertical distributions of the intensity of scattered light. The
turbulent diffusion coefficient  in the test section $D_{_{T}}$ was
of the order of  $D_{_{T}} \sim 40$ cm$^2$ / s. The turbulent
diffusion time $\tau_{_{TD}} = L_d^2 / D_{_{T}} \sim 3 $ seconds in
the scale of core flow $L_d=10$ cm. Thus the steady state particle
spatial distribution is reached during several seconds, while the
turbulence integral time scale $\tau \sim 2 \times 10^{-2}$ s. The
measurements were started several minutes after the seed was
inserted in the chamber.

\section{Turbulent thermal diffusion: theory and experiment}

Now let us discuss the effect of turbulent thermal diffusion and
compare the theoretical predictions with the experimental results.
The mean number density of particles $\bar N$ advected by a
turbulent fluid flow is given by
\begin{eqnarray}
&& {\partial \bar N \over \partial t} + {\rm div} \, [\bar N
(\bar{\bf V} + {\bf V}_{\rm eff}) - D_{_{T}} \bec{\nabla} \bar N]
= 0 \;,
\label{A2}\\
&& {\bf V}_{\rm eff} = - \tau \, \langle {\bf u}_p \, {\rm div} \,
{\bf u}_p \rangle = - D_{_{T}} (1 + \kappa) {\bec{\nabla} \bar T
\over \bar T} \;, \label{P3}
\end{eqnarray}
where $ D_{_{T}} = (\tau /3) \langle {\bf u}^2 \rangle $ is the
turbulent diffusion coefficient, $\tau$ is the momentum relaxation
time of the turbulent velocity field, ${\bf u}$ are fluctuations of
fluid velocity, $\bar{\bf V}$ is the mean fluid velocity, ${\bf
u}_p$ are fluctuations of particle velocity, $\bar T$ is the mean
fluid temperature. Coefficient $\kappa$ depends on particle inertia
(the particle size $a$), the Reynolds number and the mean fluid
temperature. In Eq.~(\ref{A2}) we neglected a small molecular mean
flux of particles caused by molecular (Brownian) diffusion,
molecular thermal diffusion (or molecular thermophoresis) and small
particle terminal fall velocity. Equation~(\ref{A2}) was previously
derived by different methods (see Elperin et al. 1996; 1997; 1998;
2000b; 2001; Pandya and Mashayek 2002; Reeks 2005).

\begin{figure}
\centering
\includegraphics[width=8cm]{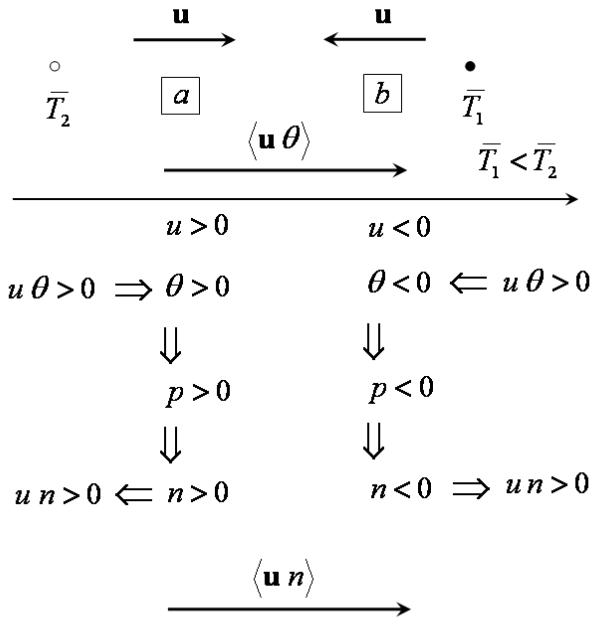}
\caption{\label{Fig5} Mechanism of turbulent thermal diffusion of
inertial particles.}
\end{figure}

For non-inertial particles, their velocity coincides with fluid
velocity $ {\bf v} =\bar{\bf V} + {\bf u},$ and ${\rm div} \, {\bf
v} \approx - ({\bf v} \cdot \bec{\nabla}) \rho / \rho \approx ({\bf
v} \cdot \bec{\nabla}) T / T ,$ where $\rho$ and $T$ are the density
and temperature of the fluid, and $\bar{\bf V} = \langle {\bf v}
\rangle $. Therefore, the effective velocity of non-inertial
particles ${\bf V}_{\rm eff} = - \tau \, \langle {\bf u} \, {\rm
div} \, {\bf u} \rangle$ is given by $ {\bf V}_{\rm eff} = -
D_{_{T}} (\bec{\nabla} \bar T) / \bar T .$ Here we used the equation
of state for an ideal gas, and neglected small gradients of the mean
fluid pressure. This effective velocity causes an additional
turbulent flux of particles directed to the minimum of mean fluid
temperature (phenomenon of turbulent thermal diffusion). Note that
turbulent thermal diffusion for non-inertial particles is the purely
kinematic effect. Indeed, the equation for the instantaneous mass
concentration, $C= m_p \, n / \rho$, of non-inertial particles in
non-isothermal flow reads
\begin{eqnarray}
{\partial C \over \partial t} + ({\bf v} \cdot \bec{\nabla}) C = {1
\over \rho} \, {\rm div} \, (D \, \rho \, \bec{\nabla} C) \;,
\label{RE1}
\end{eqnarray}
where $m_p$ is the particle mass and $n$ is the instantaneous number
density of particles. For very small molecular diffusion $D$, this
equation reads
\begin{eqnarray}
{\partial C \over \partial t} + ({\bf v} \cdot \bec{\nabla}) C
\approx 0\;, \label{RE2}
\end{eqnarray}
which implies that the mass concentration $C$ is conserved along the
fluid particle trajectory. In homogeneous turbulence all
trajectories are similar. Therefore, the number density of
non-inertial particles $n \propto \rho$, i.e., the number density of
non-inertial particles behaves locally as the fluid density. In
particular, the location of the maximum of the number density of
non-inertial particles coincides with the location of the maximum of
the fluid density, and vice versa. For very small mean fluid
pressure gradients, $\bec{\nabla} \bar \rho / \bar \rho \approx -
\bec{\nabla} \bar T / \bar T$, where $\bar \rho$ is the mean fluid
temperature. Therefore, the location of the maximum of the mean
number density of non-inertial particles coincides with the location
of the minimum of the mean fluid temperature, and vice versa. The
equation for the mean number density of non-inertial particles $\bar
N$ is equivalent to the following equation for the mean mass
concentration $ \bar C= m_p \, \bar N / \bar \rho$ of non-inertial
particles:
\begin{eqnarray}
{\partial \bar C \over \partial t} + (\bar {\bf V} \cdot
\bec{\nabla}) \bar C = {1 \over \bar \rho} \, {\rm div} \, (D_{_{T}}
\, \bar \rho \, \bec{\nabla} \bar C) \; . \label{RE5}
\end{eqnarray}
If one ignores the turbulent thermal diffusion term in
Eq.~(\ref{A2}) for the mean number density of non-inertial particles
$\bar N$, then the equation for the mean mass concentration $\bar C$
of non-inertial particles has an incorrect form.

For inertial particles, their velocity ${\bf v}_p$ depends on the
velocity of the surrounding fluid ${\bf v}$. In particular, for a
small Stokes time ${\bf v}_p \approx {\bf v} - \tau_p d{\bf v} / dt
+ {\rm O}(\tau_p^2)$ (see Maxey 1987), where $ \tau_p $ is the
Stokes time. Using the Navier-Stokes equation it can be shown that
${\rm div} \, {\bf v}_p \approx {\rm div} \, {\bf v} + \tau_p \Delta
P / \rho + {\rm O}(\tau_p^2)$ (Elperin et al. 1996). The effective
velocity~(\ref{P3}) for inertial particles is given by $ {\bf
V}_{\rm eff} = - D_{_{T}} \alpha (\bec{\nabla} \bar T) / \bar T $,
where the coefficient $\alpha = 1 + \kappa(a)$, $\, \, \kappa(a)
\propto \tau_p \propto a^2$ and  $a$ is the particle size.
Therefore, the mean particle velocity is $\bar{\bf V}_p = \bar{\bf
V} + {\bf V}_{\rm eff}$. Turbulent thermal diffusion implies an
additional non-diffusive turbulent flux of inertial particles to the
minimum of mean fluid temperature (i.e., the additional turbulent
flux of inertial particles in the direction of the turbulent heat
flux). In order to demonstrate that the directions of the turbulent
flux of inertial particles and the turbulent heat flux coincide, let
us assume that the mean temperature $\bar T_2$ at point $2$ is
larger than the mean temperature $\bar T_1$ at point $1$ (see
Fig.~5). Consider two small control volumes $"a"$ and $"b"$ located
between these two points (see Fig.~5), and let the direction of the
local turbulent velocity at the control volume $"a"$ at some instant
be the same as the direction of the turbulent heat flux $\langle
{\bf u} \theta \rangle $ (i.e., it is directed to the point $1$).
Let the local turbulent velocity at the control volume $"b"$ be
directed at this instant opposite to the turbulent heat flux (i.e.,
it is directed to the point $2$). In a fluid flow with an imposed
mean temperature gradient, pressure $p$ and velocity ${\bf u}$
fluctuations are correlated, and regions with a higher level of
pressure fluctuations have higher temperature and velocity
fluctuations. Indeed, using equation of state of the ideal gas it
can be easily shown that the fluctuations of the temperature
$\theta$ and pressure $p$ at the control volumes $"a"$ are positive,
and at the control volume $"b"$ they are negative. Therefore, the
fluctuations of the particle number density $n$ are positive in the
control volume $"a"$ (because inertial particles are locally
accumulated in the vicinity of the maximum of pressure
fluctuations), and they are negative at the control volume $"b"$
(because there is an outflow of inertial particles from regions with
a low pressure). The mean flux of particles $\langle {\bf u} n
\rangle$ is positive in the control volume $"a"$ (i.e., it is
directed to the point $1$), and it is also positive at the control
volume $"b"$ (because both fluctuations of velocity and number
density of particles are negative at the control volume $"b"$).
Therefore, the mean flux of inertial particles $\langle {\bf u} n
\rangle$ is directed, as is the turbulent heat flux $\langle {\bf u}
\theta \rangle$, towards the point~1.

\begin{figure}
\centering
\includegraphics[width=8cm]{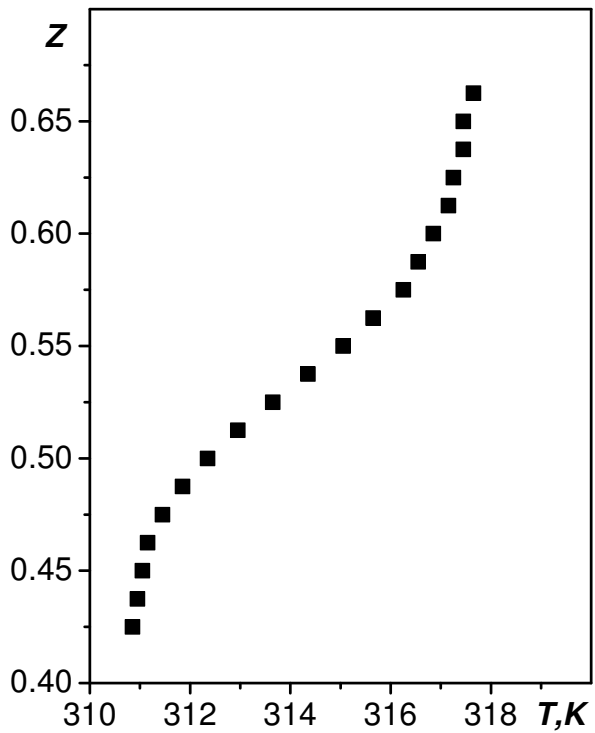}
\caption{\label{Fig6} Vertical temperature profile. Here $Z$ is a
dimensionless vertical coordinate measured in units of the height of
the chamber, and $Z=0$ at the bottom of the chamber.}
\end{figure}

The contribution of turbulent thermal diffusion to the turbulent
flux of particles is given by
\begin{eqnarray}
J_{_{T}}^{TTD} = - D_{_{T}} \alpha  {\bec{\nabla} \bar T \over \bar
T}  \bar N = - D_{_{T}}  k_{_{T}} {\bec{\nabla} \bar T \over \bar T}
\;,
\label{RE7}
\end{eqnarray}
where $ D_{_{T}} k_{_{T}} $ is the coefficient of turbulent thermal
diffusion, $k_{_{T}} = \alpha \bar N = (1+\kappa) \bar N$ is the
turbulent thermal diffusion ratio, and the coefficient $\alpha =
k_{_{T}} / \bar N $ is the specific turbulent thermal diffusion
ratio.

Neglecting the term $\bar N {\bf V}_{\rm eff}$ in Eq.~(\ref{A2})
for the mean number density of particles, we arrive at simple
diffusion equation: $\partial \bar N / \partial t = D_{_{T}}
\Delta \bar N ,$ where we neglected a small mean velocity
$\bar{\bf V}$. The steady-state solution of this equation is $\bar
N= \, $ const, i.e., a uniform spatial distribution of particles.
On the other hand, our measurements in both, the multi-fan
turbulence generator and an oscillating grids turbulence generator
(Buchholz et al. 2004; Eidelman et al. 2004) demonstrate that the
solution $\bar N= \, $ const is valid only for an isothermal
turbulent flow. Let us take into account the effect of turbulent
thermal diffusion in Eq.~(\ref{A2}). Then the steady-state
solution of Eq.~(\ref{A2}) reads: $\bec{\nabla} \bar N / \bar N =
- \alpha \bec{\nabla} \bar T / \bar T ,$ which yields
\begin{eqnarray}
{\bar N \over \bar N_0} = 1 - \alpha {\bar T - \bar T_0 \over \bar
T_0} \;, \label{R10}
\end{eqnarray}
where $\bar N_0 = \bar N(\bar T = \bar T_0)$ and $\bar T_0$ is the
reference mean temperature.

Now let us discuss the measurements of mean temperature and particle
spatial distribution in the multi-fan turbulence generator, and
compare the experimental results with the theoretical predictions.
The mean temperature vertical profile in the multi-fan turbulence
generator is shown in Fig.~6. Here $Z$ is a dimensionless vertical
coordinate measured in the units of the height of the chamber, and
$Z=0$ at the bottom of the chamber.

\begin{figure}
\centering
\includegraphics[width=8cm]{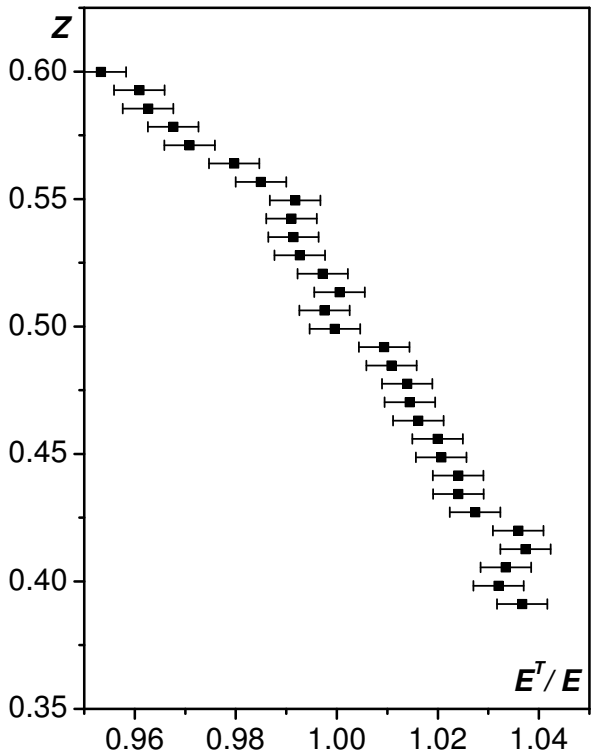}
\caption{\label{Fig7} Ratio $E^T / E$ of the normalized average
distributions of the intensity of scattered light versus the
normalized vertical coordinate $Z$.}
\end{figure}

\begin{figure}
\centering
\includegraphics[width=8cm]{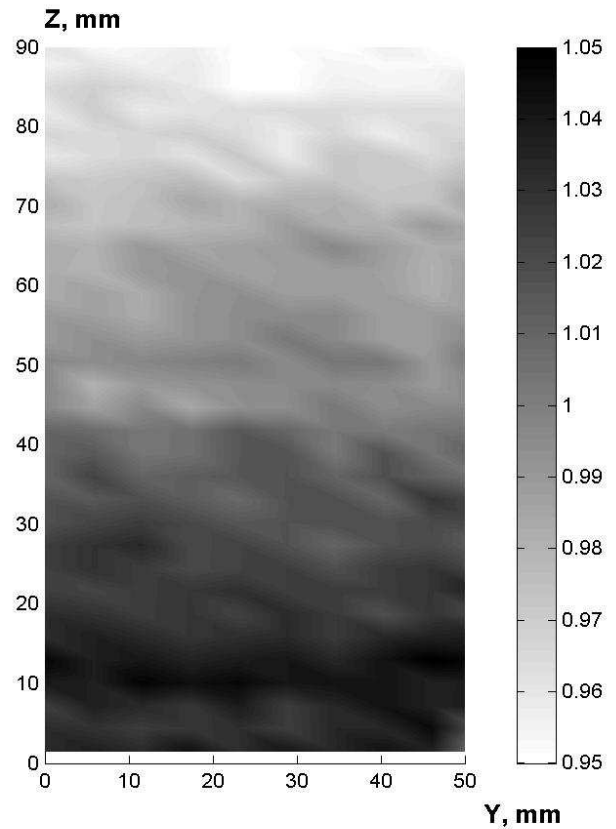}
\caption{\label{Fig8} A typical normalized $E^T/E$ image in $YZ$
plane. Here $Y$ and $Z$ are the horizontal and vertical
coordinates.}
\end{figure}

Measurements performed using different concentrations of the incense
smoke in the flow showed that the distribution of the scattered
light intensity normalized by the average over the vertical
coordinate light intensity, is independent of the mean particles
number density in the isothermal flow. In order to characterize the
spatial distribution of particle number density, $ \bar N \propto
E^T / E $, in a non-isothermal flow, the distribution of the
scattered light intensity $E$ for the isothermal case was used to
normalize the scattered light intensity $E^T$ obtained in a
non-isothermal flow under the same conditions. The scattered light
intensities $E^T$ and $E$ in each experiment were normalized by
corresponding scattered light intensities averaged over the vertical
coordinate. The ratio $E^T / E$ of the normalized average
distributions of the intensity of scattered light as a function of
the normalized vertical coordinate $Z$ is shown in Fig.~7. A typical
normalized $E^T/E$ image in $YZ$ plane is shown in Fig.~8.
Inspection of Figs. 7-8 demonstrates that particles are
redistributed in a turbulent flow with a mean temperature gradient,
e.g., they accumulate in regions with minimum mean temperature (in
the lower part of the chamber).

\begin{figure}
\centering
\includegraphics[width=8cm]{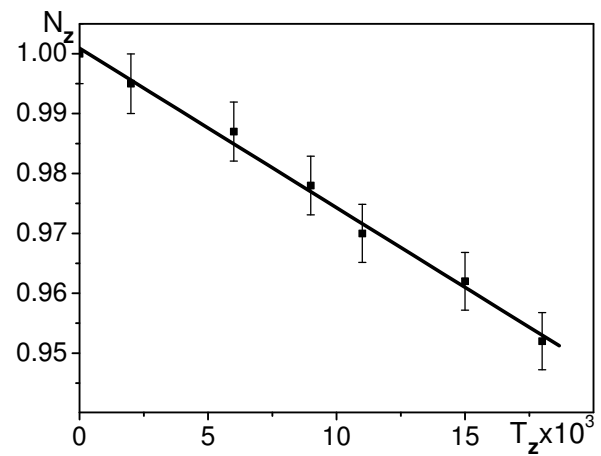}
\caption{\label{Fig9} Normalized particle number density $N_z \equiv
\bar N / \bar N_0$ versus normalized temperature difference $T_z
\equiv (\bar T - \bar T_0) / \bar T_0 $.}
\end{figure}

In order to determine the specific turbulent thermal diffusion
ratio, $\alpha$, in Fig.~9 we plotted the normalized particle number
density $N_z \equiv \bar N / \bar N_0$ versus the normalized
temperature difference $ T_z \equiv (\bar T - \bar T_0) / \bar T_0
,$ where $\bar T_0$ is the reference temperature and $ \bar N_0 =
\bar N(\bar T = \bar T_0).$ Figure~9 was plotted using the mean
temperature vertical profile shown in Fig.~6. The normalized local
mean temperatures [the relative temperature differences $ (\bar T -
\bar T_0) / \bar T_0 $] in Fig.~9 correspond to the different
locations inside the probed region. In particular, in Fig.~9 the
location of the point with reference temperature $\bar T_0$ is $Z=0$
(the lowest point of the probed region with a maximum $\bar N)$. In
these experiments we found that the coefficient $\alpha \approx
2.68$.

The specific turbulent thermal diffusion ratio $\alpha$ in the
experiments with oscillating grids turbulence  (Eidelman et al.,
2004; Buchholz et al., 2004) was $\alpha = 1.29-1.87$ (depending on
the frequency of the grid oscillations and on the direction of the
imposed vertical mean temperature gradient), while in the multi-fan
turbulence $\alpha = 2.68$. The latter value of the coefficient
$\alpha$ is larger than that obtained in the experiments in
oscillating grids turbulence, where the Reynolds numbers were
smaller than those achieved in the multi-fan turbulence generator.
Therefore, we demonstrated that the specific turbulent thermal
diffusion ratio $\alpha$ increases with increase of Reynolds number.
Note also that the experiments with oscillating grids turbulence
were performed with two directions of the imposed vertical mean
temperature gradient (for stable and unstable stratifications).

The specific turbulent thermal diffusion ratio, $\alpha = 1 +
\kappa(a)$, comprises two terms, the first one  (which equals to 1)
is independent of the particle size and the second term depends on
the size of particles. In particular, $\kappa(a) \propto \tau_p
\propto a^2$, where  $a$ is the particle size. For non-inertial
particles, $\kappa(a)=0$, and $\alpha = 1$. The deviation of the
coefficient $\alpha$ in both experiments from $\alpha=1$ is caused
by a small yet finite inertia of the particles and also by the
dependence of coefficient $\kappa$ on the Reynolds numbers. The
exact value of parameter $\alpha$ for inertial particles cannot be
found within the framework of the theory of turbulent thermal
diffusion (Elperin et al. 1996; 1997; 1998; 2000b; 2001) for the
conditions of our experiments (i.e., for large mean temperature
gradients). However, in the experiments performed for different
ranges of parameters and different directions of a mean temperature
gradient, and in two different experimental set-ups, the coefficient
$\alpha$ was more than $1$, that agrees with the theory. Therefore,
we demonstrated that turbulent thermal diffusion occurs
independently of the method of turbulence generation.

The size of the probed region did not affect our results. The
variability of the results obtained in different experiments was
within $0.5 \%$. This is caused by variability of optical
conditions, light intensity variations in a light sheet and errors
in light intensity detection. Therefore, it can be concluded that
error in particle number density measurements is less than $0.5 \%$.
Note that the contribution of the mean flow to the spatial
distribution of particles is negligibly small. In particular, the
normalized distribution of the scattered light intensity measured in
the different vertical strips in the regions where the mean flow
velocity and the coefficient of turbulent diffusion may vary, are
practically identical (the difference being only about $1 \%)$.

The effect of the gravitational settling of small particles ($0.5
- 1 \, \mu$m) is negligibly small (the terminal fall velocity of
these particles being less than $0.01$ cm/s). Due to the effect of
turbulent thermal diffusion, particles are redistributed in the
vertical direction in the chamber: particles accumulated in the
lower part of the chamber, i.e., in regions with a minimum mean
temperature. Some fraction of particles sticks to the fan
propellers and chamber walls, and the total number of particles
without feeding fresh smoke slowly decreases. The characteristic
time of this decrease is about 15 minutes. However, the spatial
distribution of the normalized number density of particles does
not change over time. It must be noted that the accuracy of the
measurements in these experiments $(\sim 0.5 \%)$ is considerably
higher than the magnitude of the observed effect $(\sim 5 \%)$.
Therefore, our experiments detected the effect of turbulent
thermal diffusion in the multi-fan turbulence generator. These
results are in compliance with the results of the previous
experiments in oscillating grids turbulence (see Buchholz et al.
2004; Eidelman et al. 2004).

\section{Conclusions}

   We studied experimentally the effect of turbulent thermal
diffusion in a multi-fan turbulence generator using Particle Image
Velocimetry and Image Processing Techniques. In a turbulent flow
with an imposed vertical mean temperature gradient (stably
stratified flow) particles accumulate in regions of minimum mean
temperature. Therefore, our experiments detected the effect of
turbulent thermal diffusion in the multi-fan turbulence generator,
i.e., non-diffusive mean flux of particles in the direction of the
mean heat flux. Turbulent thermal diffusion is an universal
phenomenon. In particular, using two very different turbulent flows
created by oscillating grids turbulence generator (Buchholz et al.
2004; Eidelman et al. 2004) and multi-fan turbulence generator, we
demonstrated that the qualitative behavior of particle spatial
distribution in non-isothermal turbulent flow is similar. The same
physics is responsible for formation of particle inhomogeneities,
i.e., competition between turbulent fluxes caused by turbulent
thermal diffusion and turbulent diffusion.

\begin{acknowledgments}
We are grateful to two anonymous referees for their very helpful and
important comments. This research was partly supported by the
German-Israeli Project Cooperation (DIP) administered by the Federal
Ministry of Education and Research (BMBF) and Israel Science
Foundation governed by the Israeli Academy of Science.
\end{acknowledgments}

\end{document}